\documentclass[conference]{IEEEtran}

\usepackage{amsmath}
\usepackage{amssymb}
\usepackage{amsfonts}
\usepackage{color}
\usepackage{graphicx}
\usepackage{algorithm}
\usepackage{algorithmic}
\usepackage{enumerate}
\usepackage{cite}
\usepackage{multirow}

\allowdisplaybreaks
\begin{document}

\title{Enhanced Quasi-Maximum Likelihood Decoding of Short LDPC Codes based on Saturation}
\author{\IEEEauthorblockN{Peng Kang$^\dag$, Yixuan Xie$^\dag$, Lei Yang$^\dag$, Chen Zheng$^\ast$, Jinhong Yuan$^\dag$, and Yuejun Wei$^\ast$}
\IEEEauthorblockA{${}^\dag$The University of New South Wales, Sydney, Australia\\}
\IEEEauthorblockA{${}^\ast$Huawei Technologies CO., LTD, Shanghai, China}}
\maketitle

\begin{abstract}
In this paper, we propose an enhanced quasi-maximum likelihood (EQML) decoder for LDPC codes with short block lengths.
After the failure of the conventional belief propagation (BP) decoding, the proposed EQML decoder selects unreliable variable nodes (VNs) and saturates their associated channel output values to generate a list of decoder input sequences.
Each decoder input sequence in the list is then decoded by the conventional BP decoder to obtain the most likely codeword.
To improve the accuracy of selecting unreliable VNs, we propose an edge-wise selection method based on the sign fluctuation of VNs' extrinsic messages.
A partial pruning stopping (PPS) rule is also presented to reduce the decoding latency.
Simulation results show that the proposed EQML decoder outperforms the conventional BP decoder and the augmented BP decoder for short LDPC codes.
It even approaches the performance of ML decoding within 0.3 dB in terms of frame error rate.
In addition, the proposed PPS rule achieves a lower decoding latency compared to the list decoding stopping rule. \color{black}
\end{abstract}

\section{Introduction}
Low-density parity-check (LDPC) codes \cite{Gallager1962LDPC} and their variations \cite{Mitchell2015scldpc,Xie2016EGSC} have found wide application areas, such as wireless communications, optical communications and storages, for their near capacity performance under low-complexity belief propagation (BP) decoding algorithms for moderate to long block lengths.
Recently, new services and applications, such as vehicle-to-everything and Internet of things, have been considered in the fifth generation mobile networks (5G) standards.
These applications require ultra reliable and low latency communications, which have drawn significant research efforts for short and low rate codes \cite{2016Durisishort,2018urllc,xiaowei2018tbcc}.
%
%

It is well-known that maximum likelihood (ML) decoder can be used to minimize the error probability of decoding under the assumption that all codewords are equiprobable.
%
%
However, its decoding complexity becomes prohibitively high when the block lengths are larger than tens of bits.
%
%
%
To reduce the decoding complexity, many researchers have investigated the suboptimal BP decoders for LDPC codes and their variations, including decoding algorithms in \cite{MacKay1999SPA}\cite{Jinghu2002Near}, and decoding architectures in \cite{Iyengar2012window}\cite{Kang2018reliability}.
However, they are commonly designed for moderate to long LDPC codes.
%
%
%
\color{black}
%

%
For short LDPC codes, several quasi-ML (QML) decoders have been investigated, e.g., \cite{Marc2001osd,Varnica2007ABP} and \cite{Scholl2016sms}.
The most common strategy adopted in these works is to introduce a \emph{reprocessing} procedure after the failure of the conventional BP decoding.
More specifically, instead of performing one-time decoding for a received codeword, a list of decoder input sequences is generated based on the received signal.
The decoder repeats the conventional BP decoding by testing each decoder input sequence in the list.
The \emph{best} candidate is chosen as the decoder output according to a certain decision metric.
%
%

The most commonly known QML decoder, namely ordered statistic decoder (OSD) \cite{Marc2001osd}, can achieve the error rate performance near that of the ML decoder.
However, the high decoding complexity introduced by matrix transformation makes the OSD unsuitable for hardware implementations.
%
%
Alternatively, other QML decoders, such as the augmented BP (ABP) decoder \cite{Varnica2007ABP} and the saturated min-sum (SMS) decoder \cite{Scholl2016sms}, select the least reliable variable nodes (VNs) and saturate their channel output values to create a list of decoder input sequences.
These sequences are reprocessed by the conventional BP decoder to obtain the most likely codeword, which is more effective for hardware implementation.
Nevertheless, the performance gap between these QML decoders and the ML decoder is still evident when the size of list is small.
%
%
%

To satisfy the new demands in 5G, e.g., the ultra-reliable low latency communications, we propose an enhanced QML (EQML) decoder based on saturation for LDPC codes with short block lengths.
We first propose a node selection method based on the sign of the VNs' extrinsic messages to improve the accuracy of the selection for unreliable VNs.
Then an efficient stopping rule based on partial pruning is presented to further reduce the decoding latency.
Simulation results demonstrate that the EQML decoder outperforms both the conventional ABP decoder and BP decoders for the LDPC codes with short block lengths and can approach the frame error rate (FER) performance of ML decoder within 0.3 dB.
Moreover, the proposed stopping rule has a lower decoding latency compared to the list decoding stopping (LDS) rule.
%

The organization of this paper is as follows. In Section \ref{conventionABP}, we briefly introduce the conventional QML decoders in the literature.
Then the proposed QML decoder is presented in Section \ref{proposeABP}, including the improved node selection method and an efficient stopping rule.
The FER performance and the decoding latency of the proposed QML decoder are investigated via simulations in Section \ref{result}.
Section \ref{conclude} draws the conclusions. \color{black}
\section{The Conventional QML Decoders}\label{conventionABP}
It is well-known that the smallest cycles in LDPC code graphs introduce a notable performance loss when the conventional BP decoders are adopted in decoding, particularly for short LDPC codes\cite{Vardy1999cycle}.
To reduce the performance gap between the conventional BP decoders and the ML decoder, the QML decoders were proposed in \cite{Varnica2007ABP} and \cite{Scholl2016sms}.
After the conventional BP decoding fails, these QML decoders execute extra operations including node selection and reprocessing, respectively.
%
%
The optimal decoded codeword is selected at the end of the reprocessing according to a certain decision metric.
The general flow of the QML decoders is described as follows.
%
%
\begin{itemize}\vspace{-1mm}
  \item Perform the conventional BP decoding. If a valid codeword is found then output the codeword. Otherwise repeat a) and b) until the stopping criterion is satisfied:
  %
  \begin{itemize}
  \item[a)] Node selection: choose unreliable VNs according to certain node selection method.
  \end{itemize}\vspace{-1mm}
  \begin{itemize}
  \item[b)] Reprocessing:
  \begin{itemize}
  \item[1)] Create a list of decoder input sequences with the channel output values saturated at the selected VN positions.
  \item[2)] Perform the conventional BP decoding with each decoder input sequence.
  \end{itemize}
  \end{itemize}\vspace{-1mm}
  \item Collect all output codewords from each reprocessing and select the optimal codeword according to the preset metric as the decoding output.\vspace{-1mm}
\end{itemize}

%
The details of the node selection and the reprocessing are presented in the remaining part of this section. \color{black}\vspace{-1mm}
\subsection{Node-wise Selection (NWS) Method}\vspace{-1mm}
Let $H$ be a parity-check matrix of size $M\times N$ for an LDPC code, where $M$ and $N$ are the number of rows and columns in $H$, respectively.
By using the graph representation, we define the set of VNs and CNs in the Tanner graph of $H$ by $\mathbb{V}$ and $\mathbb{C}$, respectively.
A node $v_n\in \mathbb{V}, 1\leq n \leq N$ is connected to a node $c_m\in \mathbb{C}, 1 \leq m \leq M$ by an edge if there is a nonzero element in the $m$-th row and $n$-th column of $H$.
Assume that $\mathbf{c} = ({c_1},{c_2}, \ldots {c_{N}})$ is the coded bits transmitted through the binary additive Gaussian noise (AWGN) channels and $\mathbf{y} = ({y_1},{y_2}, \ldots {y_{N}})$ is the received channel output.
Denoted by $\mathbf{r}(v_n)$, the initial LLR value for a node $v_n$, which can be calculated by $\mathbf{r}(v_n) = \log \frac{{P({c_n} = 0|y_n)}}{{P({c_n} = 1 | y_n)}}$.
Suppose \color{black}that $\mathbf{x} = ({x_1},{x_2}, \ldots {x}_{N})$ is the decoded codeword after the conventional BP decoding.
Let $\mathbf{s} = (s_1, s_2, \ldots s_M)$ be the syndrome vector computed from $\mathbf{x} \cdot H^T$.
The $m$-th CN is said to be unsatisfied if the $m$-th entry of the syndrome vector $\mathbf{s}$ is not equal to zero.
To describe the node selection method in \cite{Varnica2007ABP}, we define the following notations:
\begin{itemize}\vspace{-1mm}
  \item $v_s$: The selected VN for saturation.
  \item $d(v_n)$: The degree of a node $v_n\in \mathbb{V}$.
  \item $d_{\max}$: The maximum degree for a given set of VNs.
  \item $\mathbb{V}_S$: The set of VNs, which are the neighbors of all unsatisfied CNs.
  \item $\mathbb{V}_{S_{\max}}$: The set of VNs in $\mathbb{V}_S$ with the maximum degree.
\end{itemize}

As shown in \cite{Varnica2007ABP}, the node selection method of the ABP decoder is based on the reliability of all the VNs that are connected to the unsatisfied CNs.
These VNs are considered to be unreliable and are chosen for saturation with a higher priority.
Here saturation of a VN means that the initial LLR value of that VN is set to the maximum value $+ \alpha$ or the minimum value $- \alpha$. \color{black}
Note that $+ \alpha$ and $- \alpha$ refer to $+ \infty$ and $- \infty$, respectively, according to the precision or quantization of the decoder.
The ABP decoder selects a least reliable VN $v_s$ from $\mathbb{V}_S$, which is utilized to generate a list of input sequences for the reprocessing tests.
In \cite{Varnica2007ABP}, the statistical observations from simulations show that the VNs in $\mathbb{V}_{S_{\max}}$ tend to be in error with a higher probability compared to other VNs under memoryless channels.
%
%
%
The node selection method in \cite{Varnica2007ABP} is summarized in $\textbf{Algorithm \ref{alg:abpNode}}$.\vspace{-6mm}
\begin{table}[h!]
\begin{algorithm}[H]
\normalsize
\caption{The Node-wise Selection Method in \cite{Varnica2007ABP}}\label{alg:abpNode}
\begin{algorithmic}[1]
\STATE {Compute the syndrome vector $\mathbf{s} = \mathbf{x} \cdot H^T$}
\STATE {Determine the set $\mathbb{V}_S$ according to $\mathbf{s}$}
\STATE {Find ${d_{\max} = \max d(v_n), v_n \in \mathbb{V}_S, 1\leq n \leq N}$}
\STATE {Determine ${\mathbb{V}_{{S_{\max }}}} = \{ v_n \in {\mathbb{V}_S}:d(v_n) = d_{\max}\}$}
\STATE {Select ${v_s}$ as ${v_s} = \mathop {\arg \min }\limits_{v_n \in {\mathbb{V}_{{S_{\max }}}}} \left| {\mathbf{r}(v_n)} \right|$}
\end{algorithmic}
\end{algorithm}
\vspace{-8mm}
\end{table}

To prevent the selection of the already saturated VN during the reprocessing, the ABP decoder ignores $v_s$ by setting $d(v_s)=0$. \color{black}
%
%
%
In \cite{Scholl2016sms}, the proposed QML decoder chooses the saturated VNs solely based on the magnitude of $\mathbf{r}(v_n)$, which is efficient for hardware implementations.
Note that the node selection methods in both \cite{Varnica2007ABP} and \cite{Scholl2016sms} determine the saturated VNs based on the messages in node level, i.e., the node degrees or the magnitude of nodes' channel output values.
We call these kind of node selection methods as NWS methods.\vspace{-1mm}
\subsection{ABP Reprocessing}\vspace{-1mm}
After selecting the VNs, the additional decoding tests are performed by the ABP decoder.
For convenience, we define the following notations for the ABP reprocessing.
\begin{itemize}\vspace{-1mm}
  \item $j$: The current reprocessing stage.
  \item $j_{\max}$: The maximum number of reprocessing stages.
  \item $T$: The number of accomplished reprocessing tests.
  \item $\mathbb{V}^{(T)}$: The set of selected VNs for saturation after $T$ reprocessing tests.
  \item $\mathbb{M}$: The list of $2^{j}$ saturated values arranged in row vectors in stage $j$.
  \item $I_j$: The number of iterations performed by BP decoding in stage $j$.
  \item $\mathbf{x}^{(T)}$: The decoding output after $T$ reprocessing tests.
  \item $\mathcal{X}$: The set of valid codewords collected at the end of reprocessing.
  \item $\mathbf{x}_{best}$: The output codeword of the ABP decoder.
\end{itemize}
%

%
%
%
For stage $j$ reprocessing, $\mathbb{M}$ is first generated by enumerating all the possibilities for the VNs in $\mathbb{V}^{(T)}$ with the saturated values. \color{black}
Define $\mathbf{m}_t$ as the $t$-th row in $\mathbb{M}$, i.e., $\mathbf{m}_t \in \mathbb{M}, t=1,2,\ldots,2^{j}$.
%
Then the ABP decoder sets $\mathbf{r}(\mathbb{V}^{(T)})$ with different $\mathbf{m}_t$ for each reprocessing test, and performs the conventional BP decoding with the updated channel sequence as the decoder input for a fixed number of iterations $I_j$.
The above process repeats until all decoder input sequences in the list are tested and the reprocessing moves to the next stage up to $j_{\max}$.
\begin{table}
\begin{algorithm}[H]
\normalsize
\caption{The Reprocessing in \cite{Varnica2007ABP}}\label{alg:process}
\begin{algorithmic}[1]
\STATE {\textbf{Initialize:} $T=0, j=1$}
\WHILE {$j \leq j_{\max}$}
    \STATE {Determine $v_s$ according to $\textbf{Algorithm 2}$}
    \STATE {Generate the list of saturated values $\mathbb{M}$}
    \FOR {$t=1:2^{j}$}
    \STATE {Replace $\mathbf{r}(\mathbb{V}^{(T)})$ with $\mathbf{m}_t$}
    \STATE {Perform BP decoding for $I_{j}$ iterations}
    \STATE {$T = T+1$}
    \STATE {Save codeword $\mathbf{x}^{(T)}$ in $\mathcal{X}$}
    \ENDFOR
    \STATE {$j = j + 1$}
\ENDWHILE
\IF {$\mathcal{X} \neq \emptyset$}\vspace{-2mm}
    \STATE {Output ${{\mathbf{x}}_{best}} = \mathop {\mathop {\arg \min }\limits_{{{\bf{x}}^{(T)}} \in {\cal X}} \sqrt {\sum\limits_{n = 1}^N {{{({{r}(v_n)} - {{x}_n}^{(T)})}^2}} } }$}\vspace{-2mm}
\ELSE
    \STATE {Declare decoding failure}
\ENDIF
\end{algorithmic}
\end{algorithm}\vspace{-11mm}
\end{table}
%
The reprocessing of the ABP decoder is described in $\textbf{Algorithm 2}$.

%
%
Note that the reprocessing presented in \cite{Scholl2016sms} is similar to that shown in $\textbf{Algorithm 2}$.
More specific, it can be regarded as a special case of the ABP decoder in the sense that all the $j_{\max}$ VNs are selected in once.
Therefore, there are only $2^{j_{\max}}$ saturated sequences need to be tested, which reduces the decoding complexity and latency.
In addition, the choice of the parameters $j_{\max}$ and $I_j$ provides the tradeoff between decoding latency and error rate performance, which improves the flexibility of the ABP decoder to meet the requirements for various applications. \color{black}
\section{The Enhanced QML Decoder for LDPC Codes}\label{proposeABP}
In \cite{Varnica2007ABP}, it is shown that the performance of the ABP decoder can approach that of ML decoder \color{black} when the number of saturated VNs is relatively large, e.g., $j_{\max}=11$.
However, for a small or moderate $j_{max}$, there is still considerable performance loss.
To improve the error rate performance of the ABP decoder for small to moderate $j_{\max}$, we propose an edge-wise selection (EWS) method by exploiting the information about sign flips of extrinsic messages conveyed on the edges in the Tanner graph.
In addition, a partial pruning stopping (PPS) rule for the proposed QML decoder is also presented to reduce the decoding latency with negligible loss in error rate performance.
\subsection{The EWS Method}
The good performance of a QML decoder for LDPC codes depends on the accuracy of selecting the VNs for saturation.
A proper node selection method adopted in the QML decoder can notably benefit the error rate performance.
In \cite{Varnica2007ABP}, the saturated VNs are selected among the set of candidate VNs connected to a set of unsatisfied CNs.
However, a VN still has probability to be in error even though all of its neighboring CNs are satisfied.
For example, if a CN connects to even number of erroneous VNs, its associated entry in the syndrome vector can still be zero, which indicates ``undetected" errors.
Therefore, the reliability metric of each VN needs to be measured in a more precise way.

\begin{figure}[t!]
\centering\includegraphics[width=2.8in]{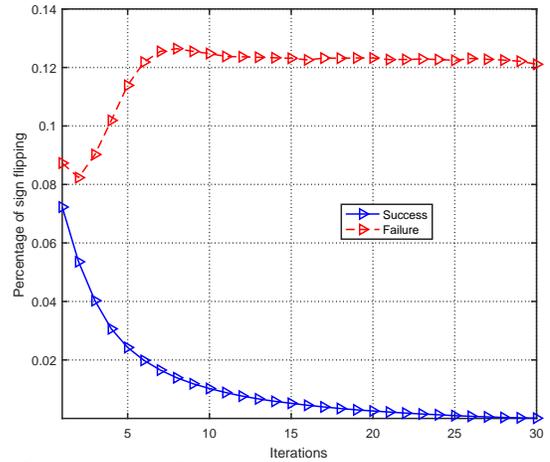}
\vspace{-4mm}
\caption{Percentage of V2C messages' sign flipping per iteration for the (96, 48) LDPC code in \cite{Scholl2016sms} with MS decoder under AWGN channels.}
\vspace{-7mm}
\label{signFlip}
\end{figure}

To improve the accuracy of the reliability measurement for VNs, we utilize the messages passed along the edges of the Tanner graph during the conventional BP decoding.
Note that during the iterative decoding process, the VNs collect the check-to-variable (C2V) messages from all the neighboring CNs and give back the extrinsic information, i.e., variable-to-check (V2C) messages, through the edges in the Tanner graph.
Fig. \ref{signFlip} shows the percentage of sign flipping of V2C messages per iteration by using a MS decoder.
We can see that the percentage of the sign flipping behaves in completely different ways for successful and unsuccessful decoding cases.
From the figure, as iteration increases, the percentage of sign flipping for V2C messages diminishes to zero in the case that the codeword can be decoded correctly.
For the case of decoding failure, the percentage of sign flipping for V2C messages saturates to a constant value after a rapid increasing in the first few iterations.
Consequently, a VN is said to be unreliable if there exists a high percentage of the sign flipping on its V2C messages.

Motivated by this observation, we utilize the sign flipping behavior on the V2C messages of each VN and propose an improved VN selection method.
Let $w_{{k,n}}$ be the number of sign flips of the V2C message passed through the $k$-th edge of the $n$-th VN.
Denoted by $w_n$, the total number of sign flips of the V2C messages on the $n$-th VN, which can be determined from \vspace{-1mm}
\begin{equation}\label{vnFluctuation}\vspace{-1mm}
w_n = \sum\nolimits_{k = 1}^{{d(v_n)}} {w_{{k,n}}}.
\end{equation}
Define $\mathbf{w}^{(T)}(v_n)$ as the number of sign flips for a node $v_n$ at the $T$-th reprocessing test, which is equal to $w_n$ computed from the $T$-th reprocessing test. \color{black}
When $T=0$, $\mathbf{w}^{(0)}(v_n)$ is initialized by the output of the conventional BP decoding.
The process of the proposed node selection method is described in \textbf{Algorithm \ref{ImproveEdge}}. 

Note that we set $\mathbf{w}^{(T)}(v_s) =0$ after the $T$-th reprocessing test to avoid selecting the same VN in different reprocessing stages. \color{black}
%
%
Compared to the NWS method, the proposed node selection method is more diverse in the sense of measuring the reliability of each VN because the selection criterion is at the edge level rather than at the node level.
Thus, we would like to call this node selection method as EWS method.
As confirmed in later simulations, the EWS method is more accurate in the sense of prioritizing the VNs that need to be saturated.
As a result, the error rate performance of the QML decoder can be improved.\vspace{-7mm}

%
\begin{figure}[t!]
\centering\includegraphics[width=2.1in]{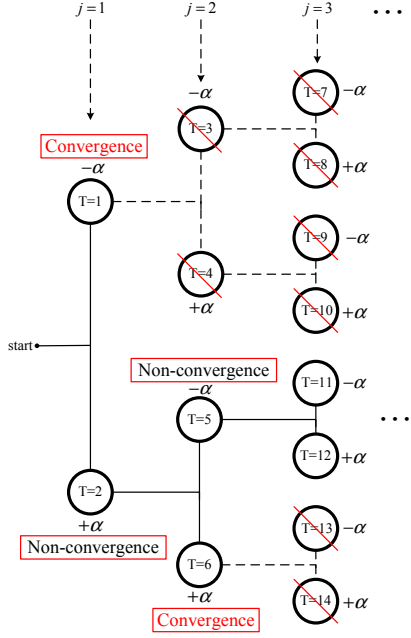}
\caption{The general branch tree of the EQML decoder operates serially in time. The order of the testing is indicated by $T$.}
\vspace{-7mm}
\label{ABP_serial}
\end{figure}
\begin{table}[h!]
\begin{algorithm}[H]
\normalsize
\caption{The EWS Method}\label{ImproveEdge}
\begin{algorithmic}[1]
\STATE {For each $v_n$, compute $w_{n}$ according to Eq. (\ref{vnFluctuation})}
\STATE {Select the VN ${v_s}$ as $v_s = \mathop {\arg \max} \limits_{v_n \in {\mathbb{V}}} {\mathbf{w}^{(T)}({v_n})}$}
    \IF{$\left| {{v_s}} \right| > 1$}
    \STATE {Select the VN with the smallest a posterior probability (APP) among $v_s$}
    \ENDIF
\end{algorithmic}
\end{algorithm}\vspace{-10mm}
\end{table}
\subsection{The PPS Rule}\vspace{-1mm}
In order to collect all possible output codewords, the ABP decoder needs to test all $2^{j_{\max}+1}-2$ sequences with the selected VNs being saturated.
This stopping rule is called ``list decoding" in \cite{Varnica2007ABP}. \color{black}
%
%
Although this LDS rule guarantees the completion of the output codeword sets, which is beneficial to the error rate performance, it causes a large decoding latency.
%
%
%
To reduce the decoding latency caused by exhaustive tests without a significant degradation in error rate performance, we propose a PPS rule to efficiently terminate the reprocessing.
Define $T_F$ as the remaining number of reprocessing tests needs to be performed, which is initialized as the total number of tests that is performed by the LDS rule, i.e., $T_F={2^{{j_{\max }} + 1}} - 2$. \color{black}
At the stage-$j$ reprocessing, the PPS rule is shown in \textbf{Algorithm \ref{ppsRule}}.
\begin{table}
\begin{algorithm}[H]
\normalsize
\caption{The PPS Rule}\label{ppsRule}
\begin{algorithmic}[1]
\IF {converge after $T$ tests}
    \STATE {Save valid codeword $\mathbf{x}^{(T)}$ in $\mathcal{X}$}
    \STATE {$T_F = T_F - 2^{j_{\max} - j}$}
    \STATE {Terminate the decoding on sub-branches thereafter}
\ELSE
    \STATE {Continue the decoding on non-convergent sub-branches.}
\ENDIF
\IF {$j = j_{\max}$ or $T_F = 0$}
\STATE {Terminate the reprocessing}
\ENDIF
\end{algorithmic}
\end{algorithm}\vspace{-11mm}
\end{table}

Fig. 2 depicts the general branch tree of an ABP decoder operates serially in time during the reprocessing, where an ABP branch is considered to be converged if there is a valid codeword found in that branch.
Compared to the LDS rule, the proposed PPS rule reduces the decoding latency by pruning the reprocessing sub-branches, which is shown as dash lines in Fig. 2.
This means that if a valid codeword is found in one branch, the reprocessing tests to be performed on the associated sub-branches thereafter are deactivated by the PPS rule. \color{black}
Note that an alternative stopping rule was proposed in \cite{Scholl2016sms}, where the decoding stops once the number of output codewords exceeds a preset threshold.
However, the threshold needs to be optimized in advance by computer simulations for different LDPC code ensembles in order to obtain the best performance.
Compared to that, the proposed PPS rule can makes a good balance between the decoding latency and the error rate performance, which is desirable for practical applications.\vspace{-4mm}
\subsection{The EQML Decoder}
Define $\mathbf{\hat{r}}^{(T)}$ as the decoder input sequence in LLR format for the $T$-th reprocessing test, which is obtained by substituting $\mathbf{r}(\mathbb{V}^{(T)})$ with $\mathbf{m}_t$.
Set the maximum number of decoding iterations as $I_{\max}$ for all BP decoding process.
By combining the EWS method and the PPS rule in the reprocessing, the proposed EQML decoder is summarized in $\textbf{Algorithm \ref{alg:qml}}$.\vspace{-4mm}
\begin{table}[h!]
\begin{algorithm}[H]
\normalsize
\caption{The EQML Decoder}\label{alg:qml}
\begin{algorithmic}[1]
\STATE Perform BP decoding with $I_{\max}$
\IF {A valid codeword is found}
\STATE {Output the codeword}
\ELSE
\STATE {\textbf{Initialize:} $T=0, j=1, T_F = 2^{j_{\max}+1}-2$}
    \WHILE {$j \leq j_{max}$}
        \STATE {Determine $v_s$ according to $\textbf{Algorithm 3}$}
        \STATE {Generate $\mathbb{M}$ by enumerating $\mathbf{r}(\mathbb{V}^{(T)})$ to $\pm \alpha$}
        \FOR {$t=1:2^{j}$}
        \STATE {Determine $\mathbf{\hat{r}}^{(T)}$ by replacing $\mathbf{r}(\mathbb{V}^{(T)})$ with $\mathbf{m}_t$}
        \STATE {Perform BP decoding with $\mathbf{\hat{r}}^{(T)}$ and $I_{\max}$}
        \STATE {$T = T+1$}
        \STATE {Perform PPS rule as in $\textbf{Algorithm 4}$}
        \ENDFOR
        \STATE {$j = j + 1$}
    \ENDWHILE
    \IF {$\mathcal{X} \neq \emptyset$}\vspace{-2mm}
        \STATE {Output ${{\mathbf{x}}_{best}} = \mathop {\mathop {\arg \min }\limits_{{{\bf{x}}^{(T)}} \in {\cal X}} \sqrt {\sum\limits_{n = 1}^N {{{({{r}(v_n)} - {{x}_n}^{(T)})}^2}} } }$}\vspace{-2mm}
    \ELSE
        \STATE {Declare decoding failure.}
    \ENDIF
\ENDIF
\end{algorithmic}
\end{algorithm}\vspace{-8mm}
\end{table}

Note that we adopt the MS decoder for all BP decoding process since it has a lower decoding complexity than the SPA decoder.
%
%
%
\color{black}
%
%
\section{numerical result}\label{result}
In this section, we compare the performance of the proposed EQML decoder with some existing decoders in terms of FER performance and decoding latency under AWGN channels.
The following LDPC codes are used in our simulations:
\begin{itemize}\vspace{-1mm}
  \item The (96,48) LDPC code: A regular LDPC code with code rate $1/2$ and code length $N=96$, is obtained from \cite{channelcodes}. Binary phase shift keying modulation is considered in the simulations and $I_{\max}$ is set to 30 for the EQML decoder.
  \item The LDPC code in 5G standard \cite{5gChannel1}: An irregular LDPC code with code rate $1/5$ and information length $K = 56$ is considered, and it has a code length $N=280$. Quadrature phase shift keying modulation is considered in the simulations and $I_{\max}$ is set to 50 for the EMQL decoder.
\end{itemize}\vspace{-2mm}
%
%
%
\subsection{The Error Rate Performance}\vspace{-1mm}
In this section, we investigate the FER performance of the proposed EQML decoder with different stopping criteria and $j_{\max}$ for the above two LDPC codes.
The FER performance of the SPA decoder and MS decoder is also shown in both Fig. 3 and Fig. 4.
Define the maximum number of iterations of the SPA decoder and the MS decoder as $I_{SPA}$ and $I_{MS}$, respectively.
For a fair comparison, we set $I_{SPA}=I_{MS}=({2^{{j_{\max }} + 1}} - 1) \cdot {I_{\max }} = (T_F+1)\cdot I_{\max }$, which is equal to the maximum number of iterations that can be used by the EQML decoder after $T$ reprocessing tests.
In addition, we show the FER performance of ML decoding in \cite{channelcodes} and the ABP decoder in \cite{Varnica2007ABP} with the LDS rule in Fig. 3.

It is shown in Fig. 3, for $j_{\max}=4$, the proposed EQML decoder with the PPS rule outperforms the MS decoder with $T_F=31$ and the SPA decoder with $T_F=31$ by about 0.5 dB, and 0.4 dB, respectively.
For $j_{\max}=6$, the gain of the EQML decoder is 0.6 dB over the SPA decoder with $T_F=127$.
The gain increases to 0.7 dB when compared to the MS decoder with $T_F=127$.
In addition, there is about 0.5 dB gain for $j_{\max}=4$ by using the proposed EQML decoder with the PPS rule comparing to the ABP decoder with LDS rule at FER=$10^{-4}$.
More importantly, for $j_{\max}=6$, the FER performance of the EQML decoder with PPS rule outperforms that of the ABP decoder with the LDS rule by about 0.7 dB and can approach that of ML decoding within 0.3 dB at FER=$10^{-4}$.
Note that the EQML decoder with PPS rule can achieve almost identical FER performance compared to that of the LDS rule for both $j_{\max}=$ 4 and 6.
%
\begin{figure}[h!]
\centering\includegraphics[width=2.9in]{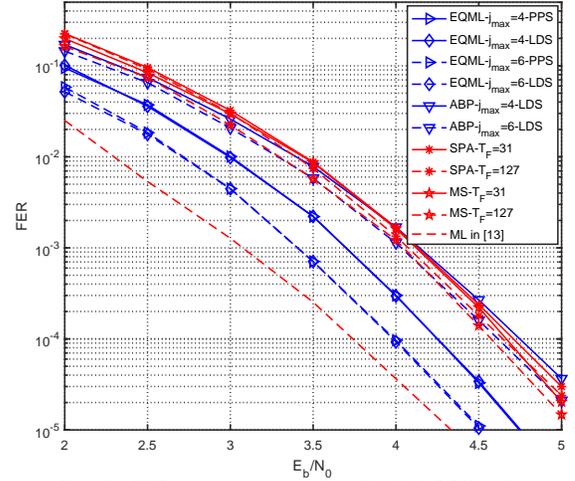}
\vspace{-4mm}
\caption{FER performance of the (96,48) LDPC code.}\vspace{-2mm}
\vspace{-1mm}
\label{EQML_(96_48)_fer}
\end{figure}
\begin{figure}[h!]
\centering\includegraphics[width=2.9in]{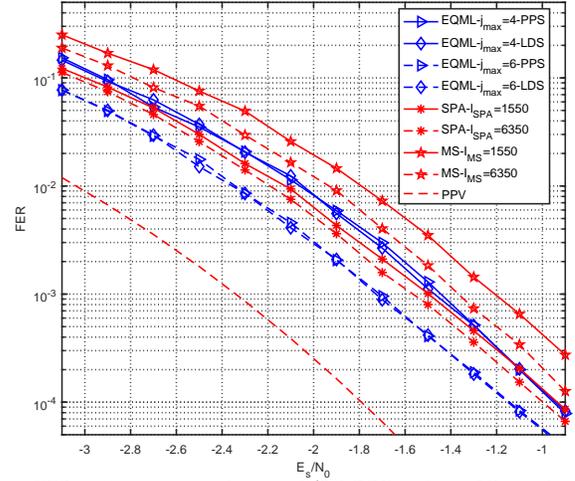}
\vspace{-4mm}
\caption{FER performance of the rate $1/5$ LDPC code in 5G standard with information length $K=56$.}
\vspace{-7mm}
\label{EQML_K56_fer}
\end{figure}

Fig. 4 depicts the FER performance of the short LDPC code in 5G standard by using various decoding methods.
For $j_{\max}=4$, it is shown that the proposed EQML decoder outperforms MS decoder by about $0.3$ dB.
When $j_{\max} = 6$, additional $0.3$ dB gain is obtained for the EQML decoder with the PPS rule in the low signal-to-noise ratio (SNR) region, i.e., $-3.1$ dB $\sim -1.9$ dB.
It demonstrates about 0.2 dB gain in FER performance when adopting the EQML decoder with the PPS rule for $j_{\max}=6$ comparing to the SPA decoder with $T_F=127$ and the gain increases to 0.4 dB when using the MS decoder with $T_F=127$.
Note that the Polyanskiy-Poor-Verd\'{u} (PPV) bound \cite{PPV} of the simulated LDPC code is also shown in Fig. 4, where the proposed EQML decoder is about 0.6 dB away from the PPV bound. \color{black}
Similarly, the EQML decoder with PPS rule can also have the similar FER performance compared to that of LDS rule for both $j_{\max}=$ 4 and 6. \vspace{-1mm}
%
%
\subsection{Decoding Latency Analysis}\vspace{-1mm}
In this section, we compare the decoding latency of the proposed EQML decoder with different stopping criteria and $j_{\max}$ for the above two LDPC codes.
Note that we only consider the EQML decoder in serial architecture, where only one decoding test runs at a time as indicated in Fig. \ref{ABP_serial}.
Therefore, we can use the average number of BP iterations used for decoding one codeword to represent the decoding latency.
Since the conventional BP decoding is executed in multiple times during the reprocessing, we define $I_{l,f}$ as the number of iterations used during the $l$-th reprocessing test of the $f$-th received codeword.
Particularly, $I_{0,f}$ refers to the number of iterations performed by the conventional BP decoding at the first time of decoding the $f$-th codeword.
To evaluate the average decoding latency of the proposed EQML decoder, let $I_{avg}$ be the average number of iterations used for decoding one codeword.
Then
\begin{equation}\label{Iavg}
{I_{avg}} = \frac{1}{F}\sum\nolimits_{f = 1}^F {({I_{0,f}} + \sum\nolimits_{l = 1}^T {{I_{l,f}}} )},
\vspace{-1mm}
\end{equation}
where $F$ represents the total number of codewords transmitted. \color{black}\vspace{-4mm}
\begin{figure}[h!]
\centering\includegraphics[width=2.9in]{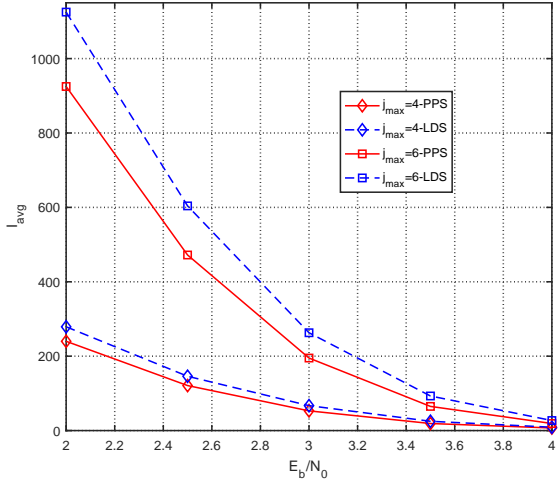}
\vspace{-1mm}
\caption{The comparison of $I_{avg}$ for the (96, 48) LDPC code with LDS and PPS rules.}
\vspace{-3mm}
\label{EQML_(96_48)_iter}
\end{figure}
\begin{figure}[h!]
\centering\includegraphics[width=2.9in]{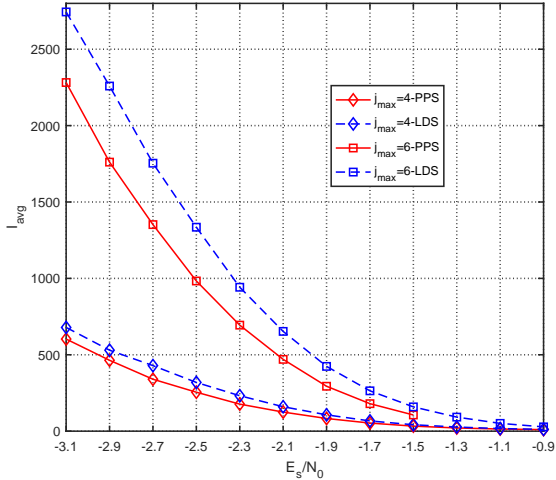}
\vspace{-2mm}
\caption{The comparison of $I_{avg}$ for the LDPC code in 5G standard with LDS and PPS rules.}
\vspace{-3mm}
\label{EQML_K56_iter}
\end{figure}

The comparison of $I_{avg}$ for the (96, 48) LDPC code decoded by the EQML decoder with the LDS and PPS rules is shown in Fig. 5.
It can be seen that our proposed EQML decoder with PPS rule requires less number of $I_{avg}$ for both $j_{\max}=4$ and $6$ compared to that of the LDS rule.
More specific, there is about 15 $\%$ less $I_{avg}$ used by PPS rule for SNR range from 2 dB to 3.5 dB.
Moreover, the reduction of the decoding latency can be even larger when $j_{\max}$ is equal to 6.
It can be seen from Fig. 5 that $I_{avg}$ used by the PPS rule is about 20 $\%$ less than that of the LDS rule for SNR range from 2 dB to 3.5 dB.

In Fig. 6, $I_{avg}$ obtained by the EQML decoder with the PPS rule for the rate-$1/5$ LDPC code from 5G standard is compared to that of the LDS rule.
We can see a similar behavior on reduction of the decoding latency as the (96, 48) LDPC code, where there is about 20 $\%$ less $I_{avg}$ required for SNR from -2.7 dB to -1.5 dB.
Note that the ABP decoder in \cite{Varnica2007ABP} can not be directly applied to the LDPC codes in 5G standard due to puncturing in the codeword.\color{black}
\section{conclusion}\label{conclude}
In this paper, we proposed an EQML decoder for decoding short LDPC codes.
In particular, we proposed an EWS method for the EQML decoder to improve the accuracy of the node selection for unreliable VNs.
%
%
Furthermore, the PPS rule was also proposed for the EQML decoder to reduce the decoding latency.
Simulation results show that the proposed EQML decoder outperforms both the conventional ABP decoder and BP decoders for the short LDPC codes and approaches the FER performance of ML decoding within 0.3 dB when $j_{\max}=6$.
Compared to the LDS rule, the proposed PPS rule can also achieve a lower decoding complexity.
\bibliographystyle{ieeetr}


%
\end{document}